\documentclass[11pt,titlepage]{article}
\usepackage[margin=1.2in]{geometry}
\usepackage{amsmath, amsbsy, amsthm, array, mathrsfs}
\usepackage{graphics}
\usepackage{physics}
\usepackage{epsfig, amssymb,latexsym,verbatim}
\usepackage{graphicx,epstopdf}
\usepackage{color}
\usepackage{dsfont, bbm}
\usepackage{relsize}
\usepackage{subcaption}
\usepackage{setspace}

\newcommand{\R}{\mathbb{R}}

\newcommand{\noin}{\noindent}
\newcommand{\bee}{\begin{eqnarray*}}
\newcommand{\ene}{\end{eqnarray*}}
\newcommand{\bec}{\begin{center}}
\newcommand{\enc}{\end{center}}
\newcommand{\be}{\begin{equation}}
\newcommand{\ee}{\end{equation}}

\newcommand{\mb}{\mathbf}
\newcommand{\bs}{\boldsymbol}
\newcommand{\tb}{\textbf}
\newcommand{\pend}{$\square$}
\newcommand{\vs}{\vskip 3mm}
\newcommand{\bi}{\begin{itemize}}
\newcommand{\ei}{\end{itemize}}

\begin{document}
\title{\LARGE Computation of least squares trimmed regression--\\ an alternative to least trimmed squares regression} 
\vs
\vs
\author{ {\sc 	Yijun Zuo and Hanwen Zuo}
	\\[2ex]
	{\small {\em Department of Statistics and Probability} and {\it Department of Computer Science
			} }\\[.5ex]
	{\small Michigan State University, East Lansing, MI 48824, USA} \\[2ex]
	{\small 
		zuo@msu.edu and zuohanwe@msu.edu}\\[6ex]
}
\date{\today}
\maketitle
\vskip 3mm
{\small
\noin	
\tb{\large Abstract}\vs
\noin
		 The least squares of depth trimmed (LST)  residuals  regression,  proposed in Zuo and Zuo (2023) \cite{ZZ23},  serves as a robust alternative to the classic least squares (LS) regression as well as a strong competitor to the famous least trimmed squares (LTS) regression of Rousseeuw (1984) \cite{R84}. Theoretical properties of the LST were thoroughly studied in \cite{ZZ23}.\vs
		
		 The article aims to promote 
		  implementation and computation of the LST residuals regression for a broad group of statisticians in the statistical practice  and demonstrates that (i)
		 the LST is as robust as the benchmark of robust regression, the LTS regression, and much more efficient than the latter.
		 (ii) It can be as efficient as (or even more efficient than) the LS in the scenario with 
		 errors that are uncorrelated with mean zero and homoscedastic with finite variance. (iii) It can be computed as fast as (or even faster than) the LTS based on a newly proposed algorithm. 

		\bigskip

\bigskip
\par

\noindent{\bf Key words and phrase:}  depth trimming,
least sum of trimmed squares of residuals, computation algorithm. 
\bigskip
\par
\noindent {\bf Running title:} least squares of trimmed residuals regression.
}
\setcounter{page}{1}		
\section{Introduction}

In the classical regression analysis, we assume that there is a relationship for a given data set $\{(\bs{x}^{\top}_i, y_i)^{\top}, i\in \{1,2, \cdots, n\}\}$: 
\be
y_i=(1,\bs{x}^{\top}_i)\bs{\beta}_0+{e}_i,~~ i\in\{1,\cdots, n\}  \label{model.eqn}
\ee
where $y_i\in \R^1$, ${\top}$ stands for the transpose, $\bs{\beta}_0=(\beta_{01}, \cdots, \beta_{0p})^{\top}$ (the true unknown parameter) in $\R^p$ and~ $\bs{x_i}=(x_{i1},\cdots, x_{i(p-1)})^{\top}$ in $\R^{p-1}$ ($p\geq 2$), $e_i\in \R^1$ is called an error term (or random fluctuation/disturbances, which is usually assumed to have zero mean and variance $\sigma^2$ in classic regression theory). That is, $\beta_{01}$ is the intercept term of the model. Write $\bs{w}_i=(1,\bs{x}'_i)^{\top}$, then one has $y_i=\bs{w}^{\top}_i\bs{\beta}_0+e_i$, which will be used interchangeably with (\ref{model.eqn}).
\vs
One wants to estimate the $\bs{\beta}_0$ based on a given sample $\mb{z}^{(n)}
:=\{(\bs{x}^{\top}_i, y_i)^{\top}, i\in\{1,\cdots, n\}\}$ from the model $y=(1,\bs{x}^{\top})\bs{\beta}_0+e$.
The difference between $y_i$ and $\bs{w^{\top}_i}{\bs{\beta}}$ is called
the ith residual, $r_i(\bs{\beta})$, for a candidate coefficient vector $\bs{\beta}$ (which is often suppressed). That is,
\be r_i:={r}_i(\bs{\beta})=y_i-\bs{w^{\top}_i}{\bs{\beta}}.\label{residual.eqn}
\ee
To estimate $\bs{\beta}_0$, the classic \emph{least squares} (LS) minimizes the sum of squares of residuals,
$$\widehat{\bs{\beta}}_{ls}=\arg\min_{\bs{\beta}\in\R^p} \sum_{i=1}^n r^2_i. $$
\vs
The LS estimator is  very popular  in practice across a broader spectrum of disciplines due to its great computability and optimal properties when the error $e_i$ follows a normal ${N}(\bs{0},\sigma^2)$ distribution.
It, however, can behave badly when the error distribution is slightly departed from the normal distribution,
particularly when the errors are heavy-tailed or contain outliers.
\vs
Robust alternatives to the $\widehat{\bs{\beta}}_{ls}$ abound in the literature. The least sum of trimmed squares (LTS) of residuals regression of Rousseeuw (1984) \cite{R84} is the benchmark of robust regression. The idea of LTS is simple, ordering the squared residuals and then trimming the larger ones and keeping at least $\lceil n/2\rceil$ squared residuals, where $\lceil ~\rceil$ is the ceiling function, the  minimizer of the sum of those {trimmed squared residuals} is called an LTS estimator:
\[
\widehat{\bs{\beta}}_{lts}:=\arg\min_{\bs{\beta}\in \R^p} \sum_{i=1}^h (r^2)_{i:n},
\]
where $(r^2)_{1:n}\leq (r^2)_{2:n}\leq \cdots, (r^2)_{n:n}$ are the ordered squared residuals and constant $h$ satisfies $\lceil n/2\rceil \leq h \leq n$.
\vs
Zuo and Zuo \cite{ZZ23} proposed the least sum of squares of depth trimmed (LST) residuals estimator, a competitor of LTS. They studied
the theoretical properties of the LST and discovered that it is not only as robust as the LTS but also much more efficient than the LTS. 
\vs
Section \ref{sec.2} introduces the LST. Section \ref{sec.3} addresses the computation issue of the LST. Section \ref{sec.4} is devoted to
the comparison of the performance of LST versus LTS and LS via concrete examples. Concluding remarks is Section \ref{sec.5} end the article. R code is given in an Appendix and a public depository https://github.com/left-github-4-codes/amlst. 
\section{Least squares of trimmed residuals regression} \label{sec.2}

\vs
\noin
\tb{Outlyingness (or depth) based trimming}~~
Depth (or outlyingness)-based trimming
scheme trims points that lie on the outskirts (i.e. points that are less deep, or outlying). The outlyingness  (or,
equivalently, depth) of a point x is defined to be  (strictly speaking, depth=1/(1+outlyingness) in \cite{Z03})
\be
O(x, \bs{x}^{(n)})=|x-\mbox{Med}(\bs{x}^{(n)})|/\mbox{MAD}(\bs{x}^{(n)}),  \label{outlyingness.eqn}
\ee
where $\bs{x}^{(n)}=\{x_1, \cdots, x_n\}$ is a data set in $\R^1$,  Med$(\bs{x}^{(n)})=\mbox{median}(\bs{x}^{(n)})$ is the median of the data points, and  MAD$(\bs{x}^{(n)})=\mbox{Med}(\{|x_i-\mbox{Med}(\bs{x}^{(n)})|,~ i\in \{1,2, \cdots, n\}\})$ is the median of absolute deviations to the center (median). It is readily seen that $O(x, \bs{x}^{(n)})$ is a generalized standard deviation, or equivalent to the one-dimensional projection depth (see \cite{ZS00} and \cite{Z03, Z06} for a high dimensional version). For notion of outlyingness, cf.  \cite{S81}, and \cite{D82, DG92}. 
\vs
\noin
\tb{Definition of the LST}~~
For a given constant $\alpha$ (hereafter assume  $\alpha\geq 1$), $\bs{\beta}$, and $\bs{z}^{(n)}$, define a set of indexes 
\be
I(\bs{\beta})=\Big\{ i:  O(r_i, \bs{r}^{(n)}) 
\leq \alpha, ~ i\in \{1, \cdots, n\} \Big\}. \label{I-beta.eqn}
\ee
where $\bs{r}^{(n)}=\{r_1, r_2, \cdots, r_n\}$ and $r_i$ is defined in (\ref{residual.eqn}).
Namely, the set of subscripts so that the outlyingness (see (\ref{outlyingness.eqn})) (or depth) of the corresponding residuals are no greater (or less) than  $\alpha$ (or $1/(1+\alpha)$). It depends on $\mb{z}^{(n)}$ 
and $\alpha$, which are suppressed  in the notation. %
For a fixed constant $\alpha$ in the depth trimming scheme, consider the quantity
\be
Q(\bs{z}^{(n)}, \bs{\beta}, \alpha):=\sum_{i=1}^{n}r_i^2\mathds{1}\left( O(r_i, \bs{r}^{(n)}\leq \alpha\right)=\sum_{i\in I(\bs{\beta})}r_i^2,\label{objective.eqn}
\ee
where $\mathds{1}(A)$ is the indicator of $A$ (i.e., it is one if A holds and zero otherwise).
Namely, residuals with their outlyingness (or equivalently reciprocal of depth minus one) greater than $\alpha$ will be trimmed.
When there is a majority ($\geq \lfloor(n+1)/2\rfloor$) identical $r_i$s, we define MAD$(\mb{r}^{(n)})=1$ (since those $r_i$  lie in the deepest position (or are the least outlying points)).
\vs
Minimizing $Q(\bs{z}^{(n)}, \bs{\beta}, \alpha)$, one gets the \emph{least} sum of \emph{squares} of {\it trimmed} (LST) residuals estimator,
\be
\widehat{\bs{\beta}}^n_{lst} :=\widehat{\bs{\beta}}_{lst}(\mb{z}^{(n)}, \alpha)=\arg\min_{\bs{\beta}\in \R^p}Q(\bs{z}^{(n)}, \bs{\beta}, \alpha).\label{lst.eqn}
\ee

It is seen that LTS essentially employs one-sided rank based trimming scheme (w.r.t. squared residuals), whereas outlyingness (or depth) based trimming is utilized in the LST.
\vs
\section{Computation and algorithm} \label{sec.3}
The following idea of computation or algorithm for the LST is based on the proof of Lemma 2.2 of \cite{ZZ23}. Assume that not all points in $\{\bs{z}_i, i \in \{1, 2, \cdots, n\}\}$ are the same and $n>2$.

\vs
\noin
\tb{Step 1} \tb{Initial step: construct $2+4p$ candidate coefficient $\bs{\beta}$s}
\bi
\item[(i)] Sample two distinct points $\bs{x}_i$ and $\bs{x}_j$  $i\neq j, i, j \in \{1, \cdots, n\}$. Assume that their $k$th components are different, that is $x_{ik}\neq x_{jk}$,  $k \in \{1, \cdots, (p-1)\}$. 
\item[(ii)] Construct two $\bs{\beta}^m$, $m\in \{0,1\}$, such that $r_i(\bs{\beta}^m)=r_j(\bs{\beta}^m)$
\[ S:=\{
\bs{\beta}^0=(0, 0, \cdots, 0, \beta_{k+1},0, \cdots, 0 ),~~
\bs{\beta}^1=(1, 0, \cdots, 0, \beta_{k+1},0, \cdots, 0 )\},
\]
where $\beta_{k+1}=(y_{i}-y_{j})/({x}_{ik}-{x}_{jk})$.
\item[(iii)] Perturbate the $l$th component of $\bs{\beta}^m$  $m\in \{0,1\}$ with an amount $\delta$, obtained $4p$ $\bs{\beta}$s
 \[ S^0:=\{ ( \beta^0_1,\cdots,
\beta^0_{l-1}, \beta^0_{l}\pm\delta, \beta^0_{l+1},\cdots, \beta^0_p)\},  
S^1:=\{ ( \beta^1_1,\cdots,
\beta^1_{l-1}, \beta^1_{l}\pm\delta, \beta^1_{l+1},\cdots, \beta^1_p)\},
\]
where $l\in \{1, \cdots, p\}$ and $\delta=0.5$ or $1$.
\vs
\ei
\noin
\tb{Step 2} \tb{Iteration step: compute the LSs based on sub-data sets}\vs

Let Betmat be a $p$ by $4p+2$ matrix storing, column-wise, all $\bs{\beta}$ in step 1 above. \\
\indent
For each $\bs{\beta}$ (column) of Betmat, do
\bi
\item[(i)] Calculate $I(\bs{\beta})$. Assume that $O(r_{i_1}, r^{(n)})\leq O(r_{i_2}, r^{(n)})\leq \cdots, \leq O(r_{i_J}, r^{(n)})$ where $J=|I(\bs{\beta})|$, the cardinality of $I(\bs{\beta})$ which is identical to the set $\{i_1,i_2, \cdots, i_J\}$.
\item[(ii)] If the inequalities in (i) are not all strict, then $\{$break; go to the next $\bs{\beta}$ in Betmat$\}$\\ else 
$\{$obtain $\widehat{\bs{\beta}}_{ls}$  and sum of squared residuals (SS$_{ls})$ based on sub-data $\{\bs{z}_i, i\in I(\bs{\beta})\}$$\}$
\item[(iii)]If (SS$_{ls}<SS_{min}$), then $\{$ $\bs{\beta}_{lst}= \widehat{\bs{\beta}}_{ls}; SS_{min}=SS_{ls}$$\}$\\
~~end do
\ei
\vspace*{-1mm}
~~~~~end for\\[1ex]
\noin
\tb{Step 3} Repeat steps 1 and 2 R times, R$\leq n(n-1)/2$, default value is one.

\vs
\noin
\tb{Output} ( $\bs{\beta}_{lst}$)
\vs
\section{Examples and comparison} \label{sec.4}
How does the LST perform? Or rather, is the LST robust, compared with the benchmark LTS? How efficient LST is, compared with LS?
Now we answer these questions by investigating the performance of the LST, comparing it with that of benchmark of robust regression, LTS, and that of benchmark of classic least squares through concrete examples.

\bec
\vspace*{-6mm}
\begin{figure}[!ht]
	\includegraphics 
	[width=0.8\textwidth] %
	{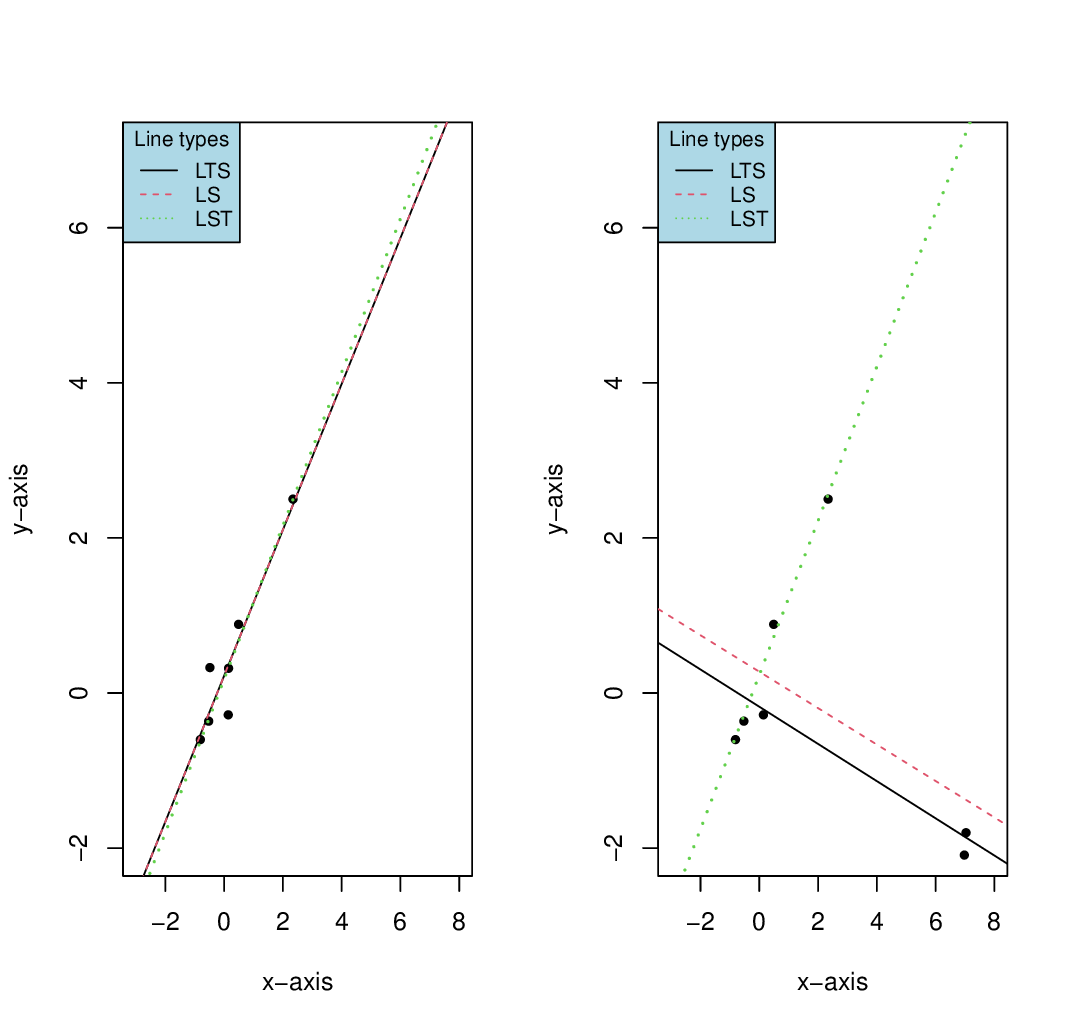}
	\caption{\footnotesize Left panel: plot of seven highly correlated normal points (with mean being the zero vector and covariance matrix with diagonal entries being one and off-diagonal entries being 0.88) and three lines given by the LST, the LTS, and the LS. 
		Right panel: The LTS line (solid black) and the LST line (dotted green), and the LS (dashed red) for the same seven highly correlated normal points but with two points contaminated nevertheless. The LS line and the LTS line are identical and drastically changed due to the two points' contamination.}
	\label{fig.one}
	\vspace*{-3mm}
\end{figure}
\enc

\vs
\noin
\tb{Example 1 (Small illustrating bivariate data sets)}. To take the advantage of graphical illustration of data sets and plots, we start with $p=2$, the simple linear regression case.\vs

\begin{figure}[!ht]
	\includegraphics 
	[width=0.8\textwidth] %
	{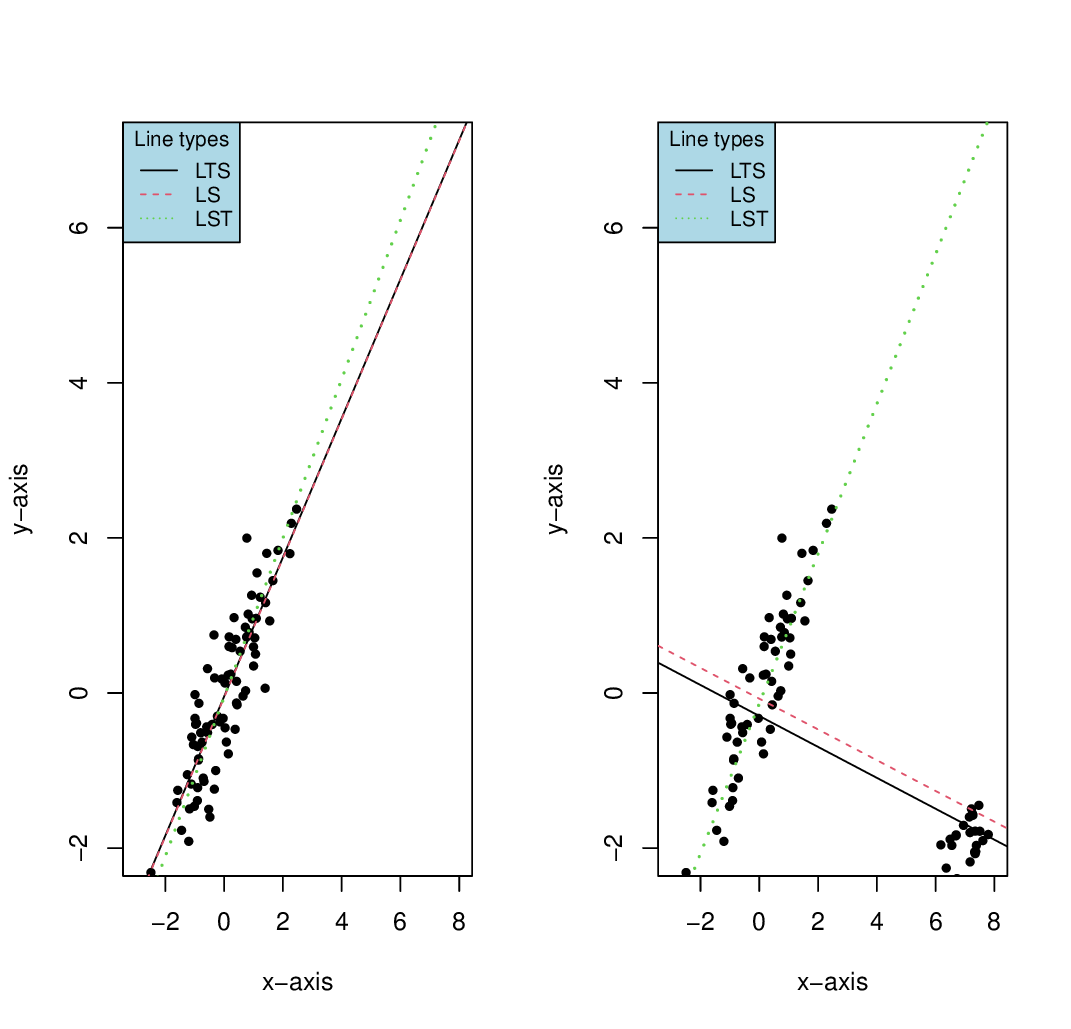}
	\caption{\footnotesize $80$ highly correlated normal points with $30\%$ of them are contaminated by other normal points. Left: scatterplot of the uncontaminated data set and three lines . Right: LTS,  LST, and LS lines. Solid black is LTS line, dotted green is the LST, and dashed red is given by the LS - parallel to LTS line in this case.}
	\label{fig.two}
	\vspace*{-8mm}
\end{figure}

We generate  seven highly correlated bivariate normal points with zero mean vector and $0.88$ as the correlation between $x$ and $y$ (the scatterplot of data points is given in the left panel of Figure \ref{fig.one}).
Two out of seven  points are adversarially contaminated in the right panel of the Figure. Three lines are fitted to the data sets. Inspecting the Figure reveals that (i) all lines
catch the overall linear pattern for perfect highly correlated normal points in left panel; (ii) when there are ($28.6\%$) contamination,
the line LTS, parallel to the LS,  is drastically attached by the two outliers. This is not the case for the line of LST, which resist the contamination (outliers). Note that, theoretically speaking, both LTS and LST can resist, asymptotically, up to $50\%$ contamination without breakdown, see \cite{RL87} and \cite{ZZ23}. One might argue that this example is not representative for LTS line since the data set is too small. To meet such concern, a similar instance is given in Figure \ref{fig.two} with sample size 
 $80$.

\bec
\vspace*{-5mm}
\enc
\vs
\noin
\tb{Example 2 (Pure Gaussian data sets)}  The classical least squares estimator, albeit is sensitive to outliers as shown above, performs the best for pure Gaussian data sets. Now we investigate the efficiency of the LTS and LST with respect to the LS in the pure Gaussian data sets scenario.\vs

For a general regression estimator $\mb{t}$,  We calculate
$\mbox{EMSE}:=\sum_{i=1}^R \|\mb{t}_i - \bs{\beta}_0\|^2/R$, the empirical mean squared error (EMSE) for $\mb{t}$. 
if $\mb{t}$ 
is  regression equivariant (see \cite{RL87} or \cite{Z21a,  Z23} for definition), then we can assume (w.l.o.g.) that the true parameter $\bs{\beta}_0=\mb{0}\in \R^p$ (see \cite{Z21b, ZZ23}, \cite{RL87}). 
Here $\mb{t}_i$ is the realization of $\mb{t}$ obtained from the ith sample with size $n$ and dimension $p$, and replication number $R$ is usually  set to be $1000$. 
\vs 
Meanwhile, we also calculate the sample variance of $\bs{t}$, by $\mbox{SVAR}(\bs{t}):=\sum_{i=1}^R \|\mb{t}_i - \bar{\bs{t}}\|^2/(R-1)$, where $\bar{\bs{t}}$ is the mean of all $\{\bs{t}_i, i\in\{1, 2, \cdots, R\}\}$, and obtain the finite sample relative efficiency (RE) of a procedure (denoted by P)  with respect to the LS by the ratio of $\mbox{RE}(P):=\mbox{SVAR}(\widehat{\bs{\beta}}_{ls})/\mbox{SVAR}(\widehat{\bs{\beta}}_{P})$. 
At the same time, we record the total time (TT) consumed by different procedures for all replications. Simulation results are listed in Table \ref{tab.one}.

\begin{table}[!h]
	\centering
	~~ Pure Gaussian data sets\\ 
	\bec
	\begin{tabular}{c c c c c cccc c}
n	&method&EMSE&SVAR&TT&RE&~~~~~EMSE&SVAR&TT&RE\\
\hline\\[-0.5ex]
&   &         &    p=5& &     &~~~  & p=10& &\\
&    LTS&0.3067&0.0696&25.985&0.5994&~~0.3017&0.1934&49.399&0.5393\\
100& LST &0.2787&0.0419&11.799&0.9954&~~0.2128&0.1042&10.596&1.0008\\
&    LS& 0.2786&0.0417&1.4166& 1.0000 &~~0.2128&0.1043&1.3384&1.0000\\[1ex]

&    LTS&0.2674&0.0300&41.750&0.6678&~~0.1825&0.0741&87.448&0.6493\\
200& LST &0.2577&0.0203&28.037&0.9869&~~0.1567&0.0482&28.490&0.9984\\
&    LS& 0.2574&0.0200&1.2762& 1.0000 &~~0.1564&0.0481&1.3128&1.0000\\[1ex]
&   &         &    p=20& &     &~~~  & p=30& &\\
&    LTS&0.5351&0.4828&159.62&0.4984&~~0.9656&0.9311&354.59&0.4468\\
100& LST &0.2919&0.2397&10.824&1.0036&~~0.4496&0.4153&11.594&1.0018\\
&    LS&0.2927&0.2406&1.3827& 1.0000 &~~0.4503&0.4161&1.4382&1.0000\\[1ex]

&    LTS&0.2219&0.1700&264.20&0.6275&~~0.3298&0.2956&572.32&0.5824\\
200& LST &0.1583&0.1063&27.969&1.0031&~~0.2060&0.1717&27.638&1.0025\\
&    LS&0.1587&0.1067&1.4514& 1.0000 &~~0.2064&0.1721&1.6345&1.0000\\
		\hline
    \end{tabular}
   \enc
\caption{\footnotesize EMSE, SVAR, TT (seconds), and  RE for LTS, LST, and LS based on $1000$ standard Gaussian samples for various $n$s and $p$s.}
\label{tab.one}
\end{table}

Inspecting the Table \ref{tab.one} reveals some stunning findings (i) the LS is by far the fastest procedure followed by LST, LTS is the slowest. It, however, is no longer the most efficient even in the pure Gaussian data points setting, the LST can be as efficient as or even more efficient than the LS. (ii) the LST is not only faster than LTS but also possesses a smaller EMSE and SVAR than LTS in all cases considered.
 As a pure R based procedure, it could speed up by an order of magnitude if employing C++, Rcpp,  or Fortran programming language just like LS or LTS does. (iii) the LTS is not efficient nor the fastest, in fact, it is the least efficient, its RE could be as low as $45\%$, which is consistent with its asymptotic efficiency that has been reported to be just $7\%$  in \cite{SHH00} or $8\%$ in \cite{MMY06} (page 132) in the literature. 
\vs\noin
\tb{Example 3 (Contaminated Gaussian data sets)} Pure Gaussian points are rare in practice. More realistic is contaminated (or mixed) Gaussian points. We now investigate the robustness of three procedures against the contamination (or outliers).
\vs
We generate $1000$ samples $\{\mb{z}_i=(\bs{x^{\top}_i},y_i)^{\top}, i\in \{1, \cdots, n\}\}$ with various $n$s  from the  normal distribution ${N}(\bs{\mu}, \bs{\Sigma})$,
where $\bs{\mu}$ is a zero-vector in $\R^p$, and $\bs{\Sigma}$ is a $p$ by $p$ matrix with diagonal entries being $1$ and off-diagonal entries being $0.9$. Then $\varepsilon\%$ of them are contaminated by $m=\lceil n\varepsilon\rceil$ points, where $\lceil\cdot\rceil$ is the ceiling function. We randomly select $m$ points of $\{\bs{z}_i$, $i\in\{1,\cdots, n\}\}$ and replace them by $(7, 7, \cdots, 7, -7)^{\top}$. We apply the three methods to the contaminated data sets, simulation results are displayed in Table \ref{tab.two}.\vs 
\begin{table}[!h]
	\centering
	~~ 
	Gaussian data sets  each with $\varepsilon $ contamination rate\\
	\bec
	\begin{tabular}{c c c c c c c c c }
method  &  EMSE  & SVAR  &TT  &RE  &~~~~~EMSE  &SVAR  &TT  &RE\\
		      \hline\\[-0.5ex]
   & p=5  &  n=50  &$\varepsilon=5\%$ & &p=5 &n=50 &$\varepsilon=10\%$ &\\
LTS&0.3948&0.1587&17.505&8.510&0.3756&0.1386&16.720&15.152\\
LST&0.3329&0.0966&7.8941&13.977&0.3363&0.0989&9.3230&21.220\\
LS &1.4223&1.3504&1.4720&1.0000&2.2726&2.0997&1.4459&1.0000\\ [1ex]
 &p=5   &n=100 & $\varepsilon=5\%$ & &  p=5   &n=100 &$\varepsilon=10\%$ & \\
LTS&0.2986&0.0616&25.368&12.033&0.2966&0.0595&25.201&17.132\\
LST&0.2798&0.0428&13.552&17.323&0.2849&0.0478&14.651&21.308\\
LS &0.8328&0.7410&1.6931&1.0000&1.1827&1.0189&1.3112&1.0000\\[1ex]
\hline\\[.5ex]
		& p=10  &n=100 & $\varepsilon=20\%$& &p=10   &n=100 &$\varepsilon=30\%$& \\
	         LTS&0.2568&0.1484&56.156&24.994&42.140&42.023&63.621&0.1083\\
            LST&0.2426&0.1343&18.208&27.629&0.2583&0.1498&22.538&30.375\\
	 	    LS &3.8447&3.7099&1.2891&1.0000&4.6918&4.5500&1.2452&1.0000\\[1ex]
         &	 p=10  &   n=200      & $\varepsilon=20\%$& &p=10&n=200 &$\varepsilon=30\%$ &\\
	         LTS&0.1725&0.0641&96.346&26.817&10.129&10.119&105.66&0.2043\\
             LST&0.1683&0.0599&46.430&28.706&0.1799&0.0715&51.175&28.927\\
		      LS &1.8479&1.7191&1.3379&1.0000&2.2121&2.0668&1.2854&1.0000\\[1ex]
\hline\\[.5ex]
		 &p=20   &n=100  & $\varepsilon=20\%$& &p=20  & n=100 &$\varepsilon=30\%$& \\
		     LTS&0.6643&0.6140&282.17&15.798&145.26&145.14&889.92&0.0864\\
            LST&0.3748&0.3229&21.219&30.040&0.4412&0.3891&24.046&32.233\\
		     LS &9.7959&9.7001&1.3300&1.0000&12.644&12.541&1.3221&1.0000\\[1ex]
 & p=20& n=200& $\varepsilon=20\%$ &&p=20  &n=200 &$\varepsilon=30\%$ &\\
		     LTS&0.1963&0.1443&403.66&29.050&31.777&31.760&850.31&0.1597\\
             LST&0.1860&0.1340&47.643&31.283&0.2109&0.1588&61.292&31.946\\
		     LS &4.2849&4.1915&1.5260&1.0000&5.1678&5.0730&1.4555&1.0000\\[1ex]
		\hline
	\end{tabular}
	\enc
	\caption{\footnotesize EMSE, SVAR, TT (seconds), and  RE for LTS, LST, and LS based on $1000$ standard Gaussian samples for various $n$s and $p$s.}
	\label{tab.two}
\end{table}

\bec
\vspace*{-8mm}
\begin{figure}[!ht]
	\includegraphics [width=14cm, height=13cm]%
	{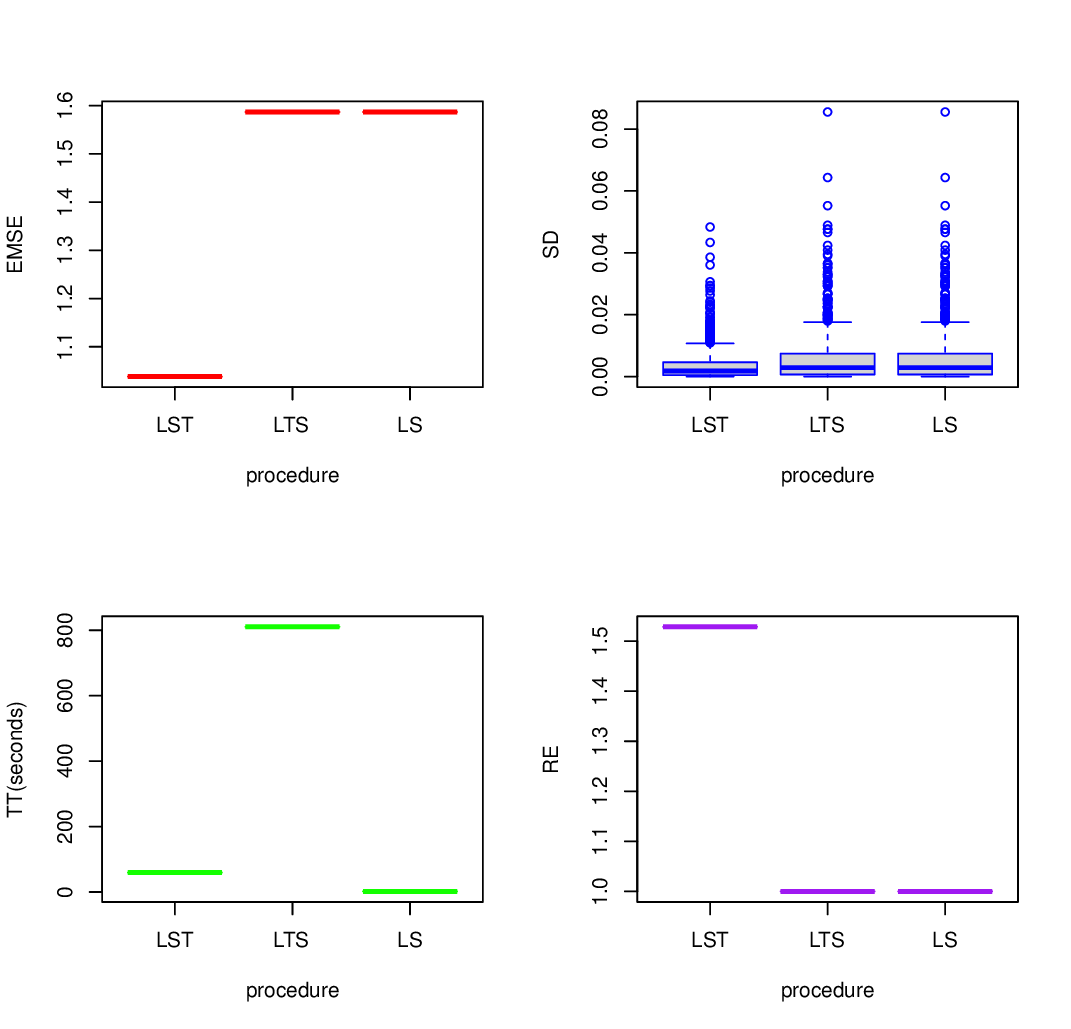}
	\caption{\footnotesize Performance of three procedures with respective $1000$ normal samples (points are highly correlated) with $p=30$ and $n=300$, each sample suffers $0\%$ contamination.}
	\label{boxplot.1}
	\vspace*{-0mm}
\end{figure}
\enc

Inspecting the Table reveals that (i) no one can run faster than LS; LST is not only always faster but also more efficient than LTS. (ii) With small contamination, LS is almost always the least efficient estimator with the exception when the contamination rate reaches  $30\%$. In the latter case, LTS shockingly becomes the least efficient. In the theory, both LTS and LST
can resist, asymptotically, up to $50\%$ contamination without breakdown (see p. 132 of \cite{RL87}, and \cite{Z21b, ZZ23}). However,
the performance of LTS is frustrating when there is $30\%$ contamination.

\vs\noin
\tb{Example 4 (\tb{Performance when $\bs{\beta}^0$ is given})}. In the calculation of EMSE above, we assumes that
	$\bs{\beta}^0=\bs{0}$ in light of regression equivariance of an estimator $\bs{t}$. In this example, we will
	provide $\bs{\beta}^0$ (for convenience write it as $\bs{\beta}_0$) and calculate $y_i$ using the formula $y_i= (1,\bs{x}^{\top}_i)\bs{\beta}^{\top}_0+e_i$, where we simulate
	$\bs{x}_i$ from a standard normal distribution with zero mean vector and identical covariance matrix. $e_i$s are i.i.d. follow a standard normal
	distribution. \vs

\begin{figure}[!ht]
	\includegraphics [width=14cm, height=13cm]%
	{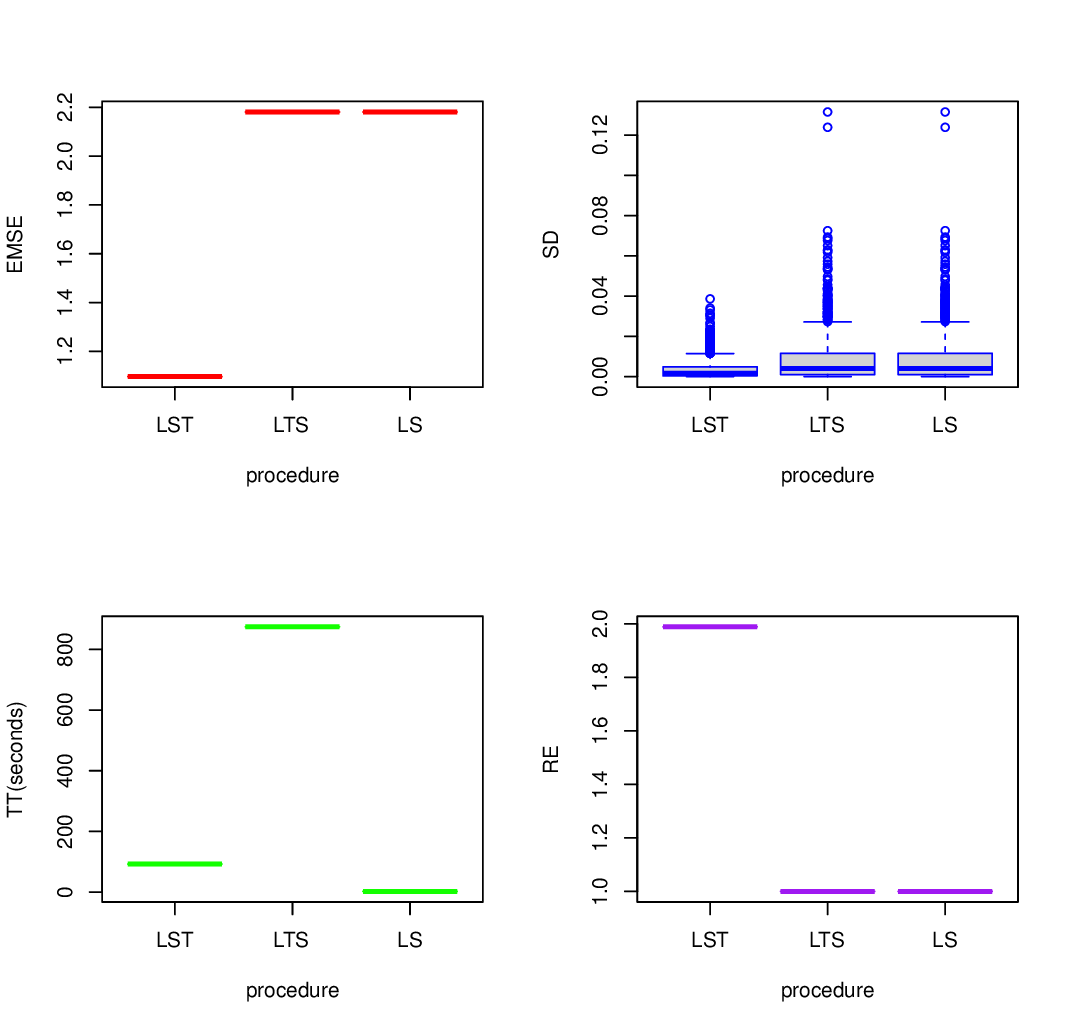}
	\caption{\footnotesize Performance of three procedures with respective $1000$ normal samples (points are highly correlated) with $p=30$ and $n=300$, each sample suffers $5\%$ contamination.}
	\label{boxplot.2}
	\vspace*{-0mm}
\end{figure}

 We set $p=30$, $n=300$ and a $\bs{\beta}_0$ with its first 15 component being $1$ and the last 15 component being $-1$. There is a $0\%$ or $5\%$ contamination for each of $1000$ normal samples (generated as in Example 3) with  the contamination scheme as: we randomly select $m=\lceil n\varepsilon\rceil$ points out of $\{\bs{z}_i$, $i\in\{1,\cdots, n\}\}$ and replace them by $(4,4, \cdots, 4, -4)^{\top}$.
We then calculate the squared deviation (SD) $(\widehat{\bs{\beta}}_i-\bs{\beta}_0)^2$ for each sample, the total time consumed by each procedures for all $1000$ samples, and the relative efficiency (the ratio of EMSE of LS versus EMSE of a procedure). The four performance criteria for different procedures are displayed graphically via boxplot in Figures \ref{boxplot.1} and \ref{boxplot.2} and numerically via Table \ref{tab.three}.
\vs
\begin{table}[!h]
	\centering $1000$ samples $\{ (\bs{x}^{\top}_i, y_i)^{\top}\}$ with
	 $y_i=(1,\bs{x}^{\top}_i)\bs{\beta}^{\top}_0+e_i$, $\bs{x}_i \sim N(0_{p\times 1}, I_{p\times p})$, $e_i\sim N(0,1)$
	\bec
	\begin{tabular}{c c c c c c}
		(n,~~p) &procedure &EMSE &SVAR &TT &RE\\
		\hline\\[-0.5ex]
		&             &              & $\varepsilon=0\%$ &&\\
		&         LTS & 1.5745       &  1.5735      &   859.56       &1.0000  \\ 
       (300, 30) &LST & 1.0346       &  1.0333      &  68.241      & 1.5218  \\
             &    LS  &1.5745       &  1.0312      &  2.0217        &1.0000 \\[1ex]
  		&             &              & $\varepsilon=5\%$ &&\\
		&         LTS & 2.1809       &  2.1791     &   874.42       &1.0000  \\ 
       (300, 30) &LST & 1.0966       &  1.0968      &  92.306      & 1.9888  \\
             &    LS  &2.1809        &  1.1066      &  2.0430        &1.0000	\\[1ex]	           
       	\hline 
	\end{tabular}
\enc
\caption{\footnotesize EMSE, SVAR, TT (seconds), and  RE for LTS, LST, and LS based on $1000$ standard Gaussian samples, $0\%$ or $5\%$ of each sample is contaminated.}
\label{tab.three}
\end{table}

\bec
\vspace*{-8mm}
\enc
\vspace*{-5mm}

\indent
Inspecting the Figures and Table reveals that (i) In terms of EMSE, LST is the best while LTS and LS are almost identically lousy. (ii)
In terms of SD (see Figures \ref{boxplot.1} and \ref{boxplot.2}), the situation is the same as the EMSE case, i.e.  LST is the best while LTS and LS are almost identically inferior. 
(iii)
In terms of computation speed, LS is by far the best and LTS is the worst. (iv) In terms of relative efficiency, LST again is the best (this is shockingly true even for $0\%$ contamination case), and  LTS and LS form the second tier.
\vs\noin
\noin
\tb{Example 5 (Performance for  large real data sets)} ~~ We consider three real data sets below. For more detailed description of data sets, see http://lib.stat.cmu.edu/datasets/\vs 
~\tb{(i) Plasma Retinol (PR) \cite{Netal89}}. This data set contains 315 observations on 14 variables. 
The goal is to investigate the relationship between personal characteristics and dietary factors, and plasma concentrations of retinol (response), beta-carotene, and other carotenoids.  
\vs
\tb{(ii) PM10} The data are a subsample of 500 observations from a data set that originate in a study where air pollution at a road is related to traffic volume and meteorological variables, collected by the Norwegian Public Roads Administration. The response variable (column 1) consist of hourly values of the logarithm of the concentration of PM10 (particles), measured at Alnabru in Oslo, Norway, between October 2001 and August 2003. The predictor variables (columns 2 to 8) are the logarithm of the number of cars per hour, temperature $2$ meter above ground (degree C), wind speed (meters/second), the temperature difference between $25$ and $2$ meters above ground (degree C), wind direction (degrees between 0 and 360), hour of day and day number from October 1. 2001. \vs
\tb{(iii) NO2} This data set is collected by the same agency at the same time period as (ii) above with the same predictor variables. The only difference is that the response variable becomes hourly values of the logarithm of the concentration of NO2 (particles).\vs
We fit the data sets with the three procedures. Since some methods depend on randomness, so we run the computation  R$=1000$ times to alleviate the randomness.
To evaluate the performance of the three procedures,
(i) we calculate the total time consumed (in seconds) for all replications for each method, (ii) the EMSE (with true $\bs{\beta}_0$ being replaced by the sample mean of R $\widehat{\bs{\beta}}$s from (i)), which then is the sample variance of all $\widehat{\bs{\beta}}$s up to a constant factor. (iii) we calculate the finite sample relative efficiency (RE) by the ratio of EMSE of LS
to EMSE of a procedure (in the numerator being zero and denominator also being zero case, RE is defined to be $100\%$ since the procedure is as efficient as the LS).
The results are reported in Table \ref{table-6}.
\vs
\begin{table}[!h]
	\centering
	\bec
	\begin{tabular}{c  c c c c c}
	 dataset & (n, ~p)	& procedure  & EMSE & TT & RE\\
	 \hline\\[-0.5ex]
	        &            & LTS  &   2241.5      &78.814 & 0\%\\
	  PR  & (315, 13)& LST       &0.0000	 &  108.09   & 100\%\\
	        &           & LS   &  0.0000    & 1.4037 &   100\%  \\[2ex]
	        &    & LTS   & 0.0011 & 152.63 & 0\%\\
PM10  & (500, 7) &LST      &0.0000 	 &143.57  & 100\%\\ 
&                & LS   &  0.0000&1.4365  &  100\%   \\[2ex]
	 &           & LTS  &       0.0017         &150.40 & 0\%\\
NO2  & (500, 7) &LST       &0.0000 	 &  141.72   & 100\%\\ 
&                & LS        &    0.0000                 &1.3968 & 100\%    \\		
	\hline		
	\end{tabular}
	\enc
	\caption{\small Total time consumed (in seconds), RE, and sample variance (EMSE) in 1000 replications by LTS, LST, and LS for three real data sets with various sample size $n$s and dimension $p$s.}
	\label{table-6}
\end{table}
Inspecting the Table reveals that (i) in terms of sample variance (or EMSE), LTS is the overall loser with the largest (non-zero) EMSE in all cases considered whereas LST and LS are the overall winner with a zero sample variance in all cases as expected (intuitively, each procedure should yield the same result for the same data set from each replication, but this is not the case for the LTS); (ii) in terms of speed, LS is the overall winner whereas LST in two out of three cases is faster than LTS; (iii) in terms of relative efficiency, LST is as efficient as the LS whereas LTS has the worst efficiency, $0\% $. 
\hfill \pend
\vs

\section{Concluding remarks} \label{sec.5}\vs

Is it possible to have a procedure that is as robust as the benchmark of robust regression, the LTS, meanwhile is more efficient 
and runs faster than the LTS? Is it possible to have a procedure that is as efficient as (or even more efficient than) the LS in the errors that are uncorrelated with mean zero and homoscedastic with finite variance scenario?  It was commonly believed that the answers to these questions were certainly impossible until this article. The latter, however, has presented promising affirmative answers.
In fact,
it has been  proved theoretically in \cite{ZZ23} that
the least squares of depth trimmed (LST) residuals regression could serve as a robust
alternative to the least squares (LS) regression and a formidable competitor to the LTS regression. This article has verified the assertion empirically via concrete examples. The computation speed of the LST could be enhanced if adopting the C++, or Rcpp, or Fortran computation language.
\vs\vs

\noin
{\textbf {\Large Supplementary Materials}}
\vs\vs\noin
The supplementary material for this article includes the following: (A) R code for LST regression (lstReg), (B) R code for examples 1, 2, 3, 4, and 5. (All are downloadable at https://github.com/left-github-4-codes/amlst).
\vs\vs
\noin
	{\textbf{\Large Acknowledgments}}
\vs \vs\noin
\noin
The author thanks Prof. Wei Shao 
for 
insightful comments and stimulus discussions.
\vs\vs

\noin
{\textbf{\Large Disclosure Statement}}
\vs\vs\noin
The author reports there are no competing interests to declare.

\vs\vs
\noin
\tb{ {\Large Funding}}\vs\vs\noin
This author declares that there is no funding received for this study.

\vs\vs
\noin
\tb{ {\Large ORCID}}\vs\vs\noin
Yijun Zuo: https://orcid.org/0000-0002-6111-3848
\vs\vs
\end{document}